\documentclass{article}
\date{
September 22, 2011%
}

\usepackage[T1]{fontenc}
\ifdefined\nopdfsync\else\usepackage{pdfsync}\fi 
\usepackage{amsmath}
\usepackage{mathrsfs}
\usepackage{booktabs}
\usepackage{bm}
\usepackage{xspace}
\usepackage{algorithmic}
\usepackage{algorithm}
\usepackage{url}
\usepackage{algorithm, algorithmic}
\floatname{algorithm}{Algorithm}

\newfloat{Algorithm}{htpb}{alg}
\newcounter{prgline}

\renewcommand{\epsilon}{\varepsilon}

\def\..{\,\mathpunct{\ldotp\ldotp}} 

\title{The Push Algorithm for Spectral Ranking} 
\author{Paolo Boldi \and Sebastiano Vigna}

\begin{document}
\bibliographystyle{plain}
\maketitle

\begin{abstract}
The push algorithm was proposed first by
Jeh and Widom~\cite{JeWSPWS} in the context of personalized PageRank
computations (albeit the name ``push algorithm'' was actually used by Andersen,
Chung and Lang in a subsequent paper~\cite{ACLUPLPG}). In this note we describe the
algorithm at a level of generality that make the computation of the spectral
ranking of any nonnegative matrix possible. Actually, the main contribution
of this note is that the description is very simple (almost trivial),
and it requires only a few elementary linear-algebra
computations. Along the way, we give new precise ways of
estimating the convergence of the algorithm, and describe some of the
contribution of the existing literature, which again turn out to be immediate
when recast in our framework.
\end{abstract}
\section{Introduction}

Let $M$ be a $n\times n$ nonnegative real matrix with entries $m_{xy}$. Without loss
of generality, we assume that $\|M\|_1=1$; this means that $M$ is \emph{substochastic} (i.e., its
row sums are at most one) and that \emph{at least one row has sum
one}.\footnote{If this is not the case, just multiply the matrix by the inverse
of the maximum row sum. The multiplication does not affect the eigenspaces, but
now the matrix satisfies the conditions above. Of course, the values of the damping
factor (see further on) have to be adjusted accordingly.}

Equivalently, we can think of the arc-weighted graph $G$ \emph{underlying} $M$.
The graph has $n$ nodes, and an arc $x\to y$ weighted by $m_{xy}$ if $m_{xy}>0$. We will
frequently switch between the matrix and the graph view, as linear matters are
better discussed in terms of $M$, but the algorithms we are interested in are
more easily discussed through $G$.

As a guiding example, given a a directed graph $G$ with $n$ nodes, $M$
can be the transition matrix of its \emph{natural walk}\footnote{We make no
assumptions on $G$, so some nodes might be dangling (i.e., without successors). In that case the corresponding rows of $M$ will be zeroed, so $M$ would 
be neither stochastic, nor a random walk in a strictly technical sense.}, whose
weights are $m_{xy}=1/d^+(x)$, where $d^+(x)$ is the outdegree of $x$
(the number of arcs going out of $x$).

We recall that the spectral radius $\rho(M)$ of $M$ coincides with the largest
(in modulus) of its eigenvalues, and satisfies\footnote{We use row vectors, so the
$\ell_1$ norm of a matrix is the maximum of the norm of the \emph{rows}.}
\[
\min_i \|M_i\|_1\leq\rho(M)\leq \max_i \|M_i\|_1 = \|M\|_1 = 1,
\]
where $M_i$ is the $i$-th row of $M$ (the second inequality is always true; the
first one only for nonnegative matrices).

Let $\bm v$ be a nonnegative vector satisfying $\|\bm v\|_1=1$ (i.e., a
distribution) and $\alpha\in[0\..1)$. The \emph{spectral
ranking}\footnote{``Spectral ranking'' is an umbrella name for techniques based
on eigenvectors and linear maps to rank entities; see~\cite{VigSR} for a
detailed history of the subject, which was studied already in the late forties.}
of $M$ with preference vector $\bm v$ and damping factor $\alpha$ is defined by
\[ \bm r = (1-\alpha)\bm v (1-\alpha M)^{-1}=(1-\alpha)\bm v
\sum_{k\geq 0}\alpha^k M^k. \] Note that the $\bm r$ needs not be a distribution,
unless $M$ is stochastic.\footnote{If $M$ is the natural walk on a graph, $\bm
r$ is not exactly PageRank~\cite{PBMPCR}, but rather the
\emph{pseudorank}~\cite{BSVPFD} associated with $\bm v$ and $\alpha$. The pseudorank is not necessarily a distribution, whereas technically
a PageRank vector always is. The distinction is however somehow
blurred in the literature, where often pseudoranks are used in place of PageRank vectors. If
$G$ has no dangling nodes, the pseudorank is exactly PageRank. Otherwise, there
are some differences depending on how dangling nodes are
patched~\cite{BPSTPTP}.} Note also that the linear operator is defined for
$\alpha\in[0\..1/\rho(M))$, but
usually estimating $\rho(M)$ is very difficult. The value $1/\rho(M)$ can
actually be attained by a limiting process which essentially makes the
damping disappear~\cite{VigSR}.

We start from the following trivial observation: while it is very difficult
to ``guess'' which is the spectral ranking $\bm r$ associated to a certain $\bm
v$, the inverse problem is trivial: given $\bm r$,
\[
\bm v = \frac1{1-\alpha}\bm r (1-\alpha M).
\]
The resulting preference vector $\bm v$ might not be, of course, a distribution
(otherwise we could obtain \emph{any} spectral ranking using a
suitable preference vector), but the equation is always true.

The observation is trivial, but its consequences are not. For instance, consider
an indicator vector $\bm\chi_x(z)=[x=z]$. If we want to obtain
$(1-\alpha)\bm\chi_x$ as spectral ranking, the associated preference
vector $\bm v$ has a particularly simple form:
\begin{equation}
\label{eq:invind}
\bm v =\frac1{1-\alpha}(1-\alpha)\bm\chi_x (1-\alpha
M)=\bm\chi_x -\alpha\sum_{x\to y}m_{xy}\bm\chi_y.
\end{equation}
We remark that in the case of a natural random walk, $m_{xy}=d(x)^{-1}$, which
does not depend on $y$ and can be taken out of the summation.
Of course, since spectral rankings are linear we can obtain
$(1-\alpha)\bm\chi_x$ multiplied by any constant just by multiplying $\bm v$ by the same costant.
 
\section{The push algorithm}

If the preference vector $\bm v$ is highly concentrated (e.g., an indicator) and
$\alpha$ is not too close to one most updates done by linear solvers or
iterative methods to compute spectral rankings are useless---either they do not
perform any update, or they update nodes whose final value will end up to be below the computational precision.

The \emph{push algorithm} uses the concentration of modifications to reduce the
computational burden. The fundamental idea appeared first in Jeh and Widom's
widely quoted paper~\cite{JeWSPWS}, albeit the notation somehow obscures the
ideas. Berkhin restated the algorithm in a different and
more readable form~\cite{BerBCAPPC}. Andersen, Chung and
Lang~\cite{ACLUPLPG} applied a specialised version of the algorithm on
symmetric graphs. All these references apply the idea to PageRank, but the
algorithm is actually an algorithm for the steady state of Markov chains with
restart~\cite{BLSGFGIP}, and it works even with substochastic matrices, so it
should be thought of as an algorithm for spectral ranking with
damping.\footnote{An implementation of the push algorithm for the computation
of PageRank is available as part of the LAW software at
\texttt{http://law.dsi.unimi.it/}.}

The basic idea is that of keeping track of vectors $\bm p$ (the current
approximation) and $\bm r$ (the \emph{residual}) satisfying
\[
\bm p + (1-\alpha)\bm r(1-\alpha M)^{-1} = (1-\alpha)\bm v(1-\alpha M)^{-1} 
\]
Initially, $\bm p=0$ and $\bm r=\bm v$, which makes the statement trivial,
but we will incrementally increase $\bm p$ (and reduce correspondingly $\bm r$).

To this purpose, we will be iteratively \emph{pushing}\footnote{The name is
taken from~\cite{ACLUPLPG}---we find it enlightening.} some node $x$. A push on
$x$ adds $(1-\alpha)r_x\bm\chi_x$ to $\bm p$. Since we must keep the
invariant true, we now have to update $\bm r$. If we think of $\bm r$ as
a preference vector, we are just trying to solve the inverse problem~(\ref{eq:invind}): by linearity, if we subtract from $\bm r$
\[
r_x\biggl(\bm\chi_x-\alpha\sum_{x\to y}m_{xy}\bm\chi_y\biggr),
\]
the value $(1-\alpha)\bm r(1-\alpha M)^{-1}$ will decrease exactly by
$(1-\alpha)r_x\bm\chi_x$, preserving the invariant.

It is not difficult to see why this choice is good: we zero an entry (the $x$-th) of $\bm r$,
and we add small positive quantities to a small (if the graph is
sparse) set of entries (those associated with the successors of $x$), increasing
the $\ell_1$ norm of $\bm p$ by $(1-\alpha)r_x$, and decreasing at least by the
same amount that of $\bm r$ (larger decreases happening on strictly
substochastic rows---e.g., dangling nodes). Note that since we do not create negative entries, it is always true that
\[
\|\bm p\|_1 + \|\bm r\|_1 \leq 1.
\]
Of course, we can easily keep track of the two norms at each
update.

The error in the estimate is  
\begin{multline*}
\bigl\|(1-\alpha)\bm r(1-\alpha M)^{-1}\bigr\|_1 =\\
(1-\alpha)\Bigl\|\bm r\sum_{k\geq0}\alpha^k M^k\Bigr\|_1 
\leq(1-\alpha)\|\bm r\|_1\sum_{k\geq0}\alpha^k \bigl\|M^k\bigr\|_1 
\leq\|\bm r\|_1.
\end{multline*}
Thus, we can control exactly the \emph{absolute additive error} of the algorithm
by controlling the $\ell_1$ norm of the residual.

It is important to notice that if $M$ is strictly substochastic
it might happen that \[\bigl\|(1-\alpha)\bm v(1-\alpha M)^{-1}\bigr\|_1<1.\]
If this happens, controlling the $\ell_1$ norm of the residual is actually
of little help, as even in the case of natural walks the norm above can be as small as
$1-\alpha$. However, since we have the guarantee that $\bm p$ is a nonnegative
vector which approximates the spectral ranking \emph{from below}, we
can simply use \[\frac{\|\bm r\|_1}{\|\bm p\|_1}\geq\frac{\|\bm r\|_1}{\bigl\|(1-\alpha)\bm v(1-\alpha M)^{-1}\bigr\|_1}\] as a measure
of relative precision, as
\[
\frac{\bigl\|(1-\alpha)\bm v (1-\alpha M)^{-1}-\bm
p\bigr\|_1}{\bigl\|(1-\alpha)\bm v(1-\alpha M)^{-1}\bigr\|_1}
=\frac{\bigl\|(1-\alpha)\bm r(1-\alpha
M)^{-1}\bigr\|_1}{\bigl\|(1-\alpha)\bm v(1-\alpha M)^{-1}\bigr\|_1}
\leq\frac{\|\bm r\|_1}{\|\bm p\|_1}.
\]

\subsection{Handling pushes}

The order in which pushes are executed can be established in many different
ways. Certainly, to guarantee relative error $\epsilon$ we need only push nodes
$v$ such that $r_x>\epsilon \|\bm p\|_1/n$, as if all nodes fail to satisfy the
inequality then $\|\bm r\|_1/\|\bm
p\|_1\leq\epsilon$.

The obvious approach is that of keeping an indirect priority queue (i.e., a queue in
which the priority of every element can be updated at any time) containing the
nodes satisfying the criterion above (initially, just the support of $\bm v$)
and returning them in order of decreasing $r_x$. Nodes are added to the queue
when their residual is larger than $\epsilon \|\bm p\|_1/n$. Every time a
push is performed, the residual of successors of the pushed node are updated and
the queue is notified of the changes.

While this generates potentially an $O(\log n)$ cost per arc visited (to
adjust the queue), in intended applications the queue is always very small, and
pushing larger values leads to a faster decrease of $\|\bm r\|_1$. 


An alternative approach is to use a FIFO queue (with the proviso that nodes
already in the queue are not enqueued again). In this case, pushes are not
necessarily executed in the best possible order, but the queue 
has constant-time access. 

Some preliminary experiments show that the two approaches are
complementary, in the sense that in situations where the number of nodes in the
queue is relatively small, a priority code reduces significantly the number of
pushes, resulting in a faster computation. However, if the queue becomes large
(e.g., because the damping factor is close to one), the logarithmic burden at
each modification becomes tangible, and using a FIFO queue yields a faster computation in spite of the higher number of pushes.

In any case, to reduce the memory footprint for large graphs it is essential
to keep track of the bijection between the set of visited nodes and an
identifier assigned incrementally in discovery order. In this way, all vectors
involved in the computation can be indexed by discovery order, making their
size dependent just on the size of the visited neighbourhood, and not
on the size of the graph.

\subsection{Convergence}

There are no published results of convergence for the push algorithm. Andersen,
Chung and Lang provide a bound not for convergence to the pseudorank, but rather
for convergence to the ratio between the pseudorank and the stationary state of
$M$ (which in their case---symmetric graphs---is trivial, as it is proportional
to the degree).

In case a priority queue is used to select the nodes to be pushed, when the
preference vector is an indicator $\bm\chi_x$ the amount of rank going to $\bm p$ at the first step is exactly $1-\alpha$. In the
follow $d(x)$ steps, we will visit either the successors of $x$, whose
residual is $\alpha/d(x)$, or some node with a larger residual, due to
the prioritization in the queue. As a result, the amount of rank going to $\bm p$
will be at least $\alpha(1-\alpha)$. In general, if 
$P_x(t)$ is the \emph{path function} of $x$ (i.e., $P_x(t)$ is the
number of paths of length at most $t$ starting from $x$), after
$P_x(t)$ pushes the $\ell_1$ norm of $\bm r$ will be at most
$1-(1-\alpha)\sum_{0\leq k\leq t}\alpha^k=\alpha^{t+1}$.

\subsection{Some remarks}

\paragraph{Precomputing spectral rankings.} 

Another interesting remark\footnote{Actually, a translation of
Jeh and Widom's approach based on \emph{partial vectors}, which was
restated by Berkhin's under the name \emph{hub decompositions}~\cite{BerBCAPPC}.
Both become immediate in our setting.} is that if during the
computation we have to perform a push on a node $x$ \emph{and we happen to know
the spectral ranking of $x$} (i.e., the spectral ranking with
preference vector $\bm\chi_x$) we can simply zero $r_x$ and add the
spectral ranking of $x$ multiplied by the current value of $r_x$ to
$\bm p$. Actually, we could even \emph{never push $x$} and just add the
spectral ranking of $x$ multiplied by $r_x$ to $\bm p$ at the end of
the computation.

Let us try to make this observation more general. Consider a set $H$ of
vertices whose spectral ranking is known; in other words, for each $x
\in H$ the vector
\[
\bm s_x = (1-\alpha)\bm \chi_x (1-\alpha M)^{-1}
\]
is somehow available.
At every step of the algorithm, the invariant equation
\[
\bm p + (1-\alpha)\bm r(1-\alpha M)^{-1} = (1-\alpha)\bm v(1-\alpha M)^{-1} 
\]
can be rewritten as follows: let $\bm r'$ be the vector obtained from $\bm r$
after zeroing all entries outside of $H$, and let $\bm p'=\bm p+\sum_{x \in H} r_x \bm s_x$.
Then clearly
\[
\bm p' + (1-\alpha)\bm r'(1-\alpha M)^{-1} = (1-\alpha)\bm v(1-\alpha M)^{-1}.
\]
Note that
\[
	\|\bm r'\|_1=\sum_{x \not\in H} r_x
\]
and
\[
	\|\bm p'\|_1=\|\bm p\|_1 + \sum_{x \in H} r_x \cdot \|\bm s_x\|_1.
\]

So we can actually execute the push algorithm keeping track of $\bm p$ and $\bm r$ but
considering (virtually) to possess $\bm p'$ and $\bm r'$, instead; to this aim, we proceed as follows:
\begin{itemize}
	\item we never add nodes in $H$ to the queue;
	\item for convergence, we consider the norms of $\bm p'$ and $\bm r'$, as computed above;
	\item at termination, we adjust $\bm p$ obtaining $\bm p'$ explicitly.
\end{itemize}

Berkhin~\cite{BerBCAPPC} notes that when computing the spectral ranking of $x$
we can use $x$ as a hub after the first push. That is, after the first push we
will never enqueue $x$ again. At the end of the computation, we simply multiply
the resulting spectral ranking by $1+r_x+r_x^2+\cdots = 1/(1-r_x)$. In this
case, $\|\bm p\|_1$ must be divided by $1-r_x$ to have a better estimate of the
actual norm. Preliminary experiments on web and social graphs show that the
reduction of the number of pushes is very marginal, though.

\paragraph{Patching dangling nodes.}
Suppose that, analogously to what is usually done in power-method computations,
we may patch dangling nodes. More precisely, suppose that
we start from a matrix $M$ that has some zero rows (e.g., the natural walk of a
graph $G$ with dangling nodes), and then we obtain a new matrix $P$ (for ``patched'') by \emph{substituting} each zero row with some distribution $\bm u$, as yet unspecified.

It is known that avoiding at all the patch is equivalent to using
$\bm u=\bm v$~\cite{BSVPFD}, modulo a scale
factor that is computable starting from the spectral ranking itself.
More generally, if $\bm u$ coincides with the distribution that is being used for preference, no patching
is needed provided that the final result is normalized.

For the general case (where $\bm u$ may not coincide with $\bm v$), we
can adapt the push method described above as follows:  we keep track of vectors $\bm p$ and $\bm r$ and of a scalar
$\theta$ representing the amount of rank that went through dangling nodes. The
equation now is
\[
\bm p + (1-\alpha)(\bm r +\theta\bm u)(1-\alpha P)^{-1}
=(1-\alpha)\bm\chi_x(1-\alpha P)^{-1}
\]

When $\bm p$ is increased by $(1-\alpha) r_x \bm\chi_x$, we have to modify 
$\bm r$ and $\theta$ as follows:
	\begin{itemize}
		\item if $x$ is not dangling, we subtract from $\bm r$ the vector
		\[
			r_x \biggl( \bm\chi_x - \alpha \sum_{x\to y} m_{xy}
			\bm\chi_y\biggr);
		\]
		\item if $x$ is dangling, we subtract just 
		\[
			r_x \bm\chi_x 
		\]
		and \emph{increase} $\theta$ by $\alpha r_x$.
	\end{itemize}
At every computation step the approximation of the spectral ranking
will be given by $\bm p'=\bm p+\theta\bm s$, where $\bm s$ is the spectral
ranking of $P$ with preference vector $\bm u$ and damping factor $\alpha$.\footnote{Of course, $\bm s$ must be precomputed using any standard
method. If $M$ is the natural walk of a graph $G$, this is exactly the PageRank
vector for $G$ with preference vector $\bm u$.} As on the case of hubs, we
should consider $\|\bm p'\|_1=\|\bm p\|_1+\theta\|\bm s\|_1$ when establishing convergence.

\hyphenation{ Vi-gna Sa-ba-di-ni Kath-ryn Ker-n-i-ghan Krom-mes Lar-ra-bee
  Pat-rick Port-able Post-Script Pren-tice Rich-ard Richt-er Ro-bert Sha-mos
  Spring-er The-o-dore Uz-ga-lis }


\begin{thebibliography}{1}

\bibitem{ACLUPLPG}
Reid Andersen, Fan R.~K. Chung, and Kevin~J. Lang.
\newblock Using {P}age{R}ank to locally partition a graph.
\newblock {\em Internet Math.}, 4(1):35--64, 2007.

\bibitem{BerBCAPPC}
Pavel Berkhin.
\newblock Bookmark-coloring approach to personalized {P}age{R}ank computing.
\newblock {\em Internet Math.}, 3(1), 2006.

\bibitem{BLSGFGIP}
Paolo Boldi, Violetta Lonati, Massimo Santini, and Sebastiano Vigna.
\newblock Graph fibrations, graph isomorphism, and {P}age{R}ank.
\newblock {\em RAIRO Inform. Th{\'e}or.}, 40:227--253, 2006.

\bibitem{BPSTPTP}
Paolo Boldi, Roberto Posenato, Massimo Santini, and Sebastiano Vigna.
\newblock Traps and pitfalls of topic-biased {P}age{R}ank.
\newblock In William Aiello, Andrei Broder, Jeannette Janssen, and Evangelos
  Milios, editors, {\em WAW 2006. Fourth Workshop on Algorithms and Models for
  the Web-Graph}, number 4936 in Lecture Notes in Computer Science, pages
  107--116. Springer--Verlag, 2008.

\bibitem{BSVPFD}
Paolo Boldi, Massimo Santini, and Sebastiano Vigna.
\newblock Page{R}ank: {F}unctional dependencies.
\newblock {\em ACM Trans. Inf. Sys.}, 27(4):1--23, 2009.

\bibitem{JeWSPWS}
Glen Jeh and Jennifer Widom.
\newblock Scaling personalized web search.
\newblock In {\em Proc. of the Twelfth International World Wide Web
  Conference}, pages 271--279. ACM Press, 2003.

\bibitem{PBMPCR}
Lawrence Page, Sergey Brin, Rajeev Motwani, and Terry Winograd.
\newblock The {P}age{R}ank citation ranking: Bringing order to the web.
\newblock Technical report, Stanford Digital Library Technologies Project,
  Stanford University, Stanford, CA, USA, 1998.

\bibitem{VigSR}
Sebastiano Vigna.
\newblock Spectral ranking, 2009.

\end{thebibliography}
\end{document}